\documentclass[12pt,preprint]{aastex}
\usepackage{amssymb}
\usepackage{psfig}

\newcommand\flux{erg sec$^{-1}$cm$^{-2}$}

\newcommand\ibu{\indent$\bullet$ }

\newcommand\cha{{\it Chandra} }

\received{2003 September 22}
\begin{document}
\title{Long term X-ray spectral variability of the nucleus of M~81}
\author{ V. La Parola\altaffilmark{1,2}, G. Fabbiano\altaffilmark{2}, M.
Elvis\altaffilmark{2}, F. Nicastro\altaffilmark{2}, D.W.
Kim\altaffilmark{2}, G. Peres\altaffilmark{1}} 
\affil{$^{1}$ Dipartimento di Scienze Fisiche ed Astronomiche, 
Sezione di Astronomia, Piazza del Parlamento 1, 90134 Palermo, Italy; }
\affil{$^{3}$ Harvard Smithsonian Center for Astrophysics, 60 Garden St.,
02138 Cambridge, MA} 
\altaffiltext{2}{e-mail: laparola@oapa.astropa.unipa.it}

\begin{abstract}
We have analysed the soft X-ray emission from the nuclear source of the nearby
spiral galaxy M~81, using the available data collected with
ROSAT, ASCA, BeppoSAX and \cha. The source flux is highly variable, showing (sometimes
dramatic: a factor of 4 in 20 days) variability at different timescales, from 2
days to 4 years, and in particular a steady 
increase of the flux by a factor of $\gtrsim 2$ over 4 years, broken by rapid 
flares. After accounting for the extended component resolved by \cha, the
nuclear soft X-ray spectrum (from ROSAT/PSPC, BeppoSAX/LECS and \cha data)
cannot be fitted well with a single absorbed power-law model. Acceptable fits
are obtained adding an extra component, either a multi-color black body  (MCBB)
or an absorption feature. In the MCBB case the inner accretion disk would be far 
smaller than the Schwartzchild radius for the $3-60\times 10^6M_{\odot}$
nucleus requiring a strictly edge-on inclination of the disk, even if the 
nucleus is a
rotating Kerr black hole. The temperature is 0.27 keV, larger than expected from
the accretion disk of a Schwartzchild black hole, but consistent
with that expected from a Kerr black hole. 
In the power-law + absorption feature
model we have either high velocity (0.3 $c$)  infalling C{\sc v} clouds or
neutral C{\sc i} absorption at rest. In both cases the C:O overabundance is a
factor of 10.
\end{abstract}

%\today

\section{INTRODUCTION}
\label{intro}
M81 (NGC 3031) is a  nearby galaxy \citep[3.6 Mpc,][]{fre}, with well defined 
spiral arms and a prominent bulge, with a structure similar to that of M31. Its 
nucleus is the nearest example of a low luminosity AGN \citep{pei,elvis} and has 
been studied at all frequencies; its emission properties make it both a 
LINER \citep[Low Ionization Nuclear Emission line Region;][]{ho} and a LLAGN 
(Low Luminosity AGN): it contains a broad component of the $H\alpha$ emission 
line \citep{pei}, a compact 
radio core \citep{biet} and a variable point-like X-ray source ($L_X \sim 10^{40}$
erg/s), M81~X-5 \citep{elvis,fab88} with a power law 
continuum with photon index $\Gamma\sim1.85$ in the 2-10 keV 
energy band \citep{ishi,pelle}. Broad line velocity \citep{ho} and stellar 
dynamics \citep{bower} studies  suggests a super-massive black hole of 
$3\times 10^6M_{\odot}< M_{BH}<6\times 10^7M_{\odot}$. 
%einstein flux threshold: $\sim 2\times 10^{-12}$
        
ASCA (0.5-10.0~keV, \citealp{ishi}) and BeppoSAX (0.1-100 keV, 
\citealp{pelle}) spectral analyses of the nuclear X-ray source show, in 
addition to the power-law component ($\Gamma \sim 1.85$), a 6.7 keV He-like Fe 
resonance line and a thermal component with $kT\sim 0.86$ keV \citep{ishi}.
A more recent study by \citet{imm}, which uses all the co-added available PSPC 
observations and fixes the power-law photon index to the ASCA/BeppoSAX
$\Gamma=1.85$ confirms this soft component and models it with a two-temperature 
optically thin plasma, ascribing it to the presence of optically thin diffuse 
gas (kT=0.15 keV) and to a population of X-ray binaries and SNR (kT=0.63 keV).
However this type of source would be harder than 0.63 keV.

Significant variability in the flux of the nuclear X-ray source on a short 
timescale ($\sim600$~s) was reported by \citet{barr} using EXOSAT data. 
Longer timescale ($\sim2$ days to few years) variability has also been reported
\citep{pelle,ishi,imm}. No spectral variability has been reported;
however, a full study of the spectral behaviour using the wealth of data 
collected from the nucleus of M81 has not been done so far. 
In the present paper we revisit the soft X-ray emission of the 
nucleus of M81 through the observations of ROSAT instruments concentrating on
variability, and compare these data with those from ASCA, BeppoSAX and \cha. 
In particular, the ROSAT/PSPC data give us the opportunity to study any 
spectral variation of soft X-ray emission that occurred during the extensive 
ROSAT coverage of M81. This paper is structured as follows:
Section~\ref{data} illustrates the data and their reduction; in
Section~\ref{time} we describe the source variability;
Section~\ref{spec} is devoted to spectral analysis; our results are
discussed in Section~\ref{disc}.

\section{OBSERVATIONS AND DATA REDUCTION}
\label{data}
A log of all the observations of M81 is given in Table~\ref{log}. 
M81 was observed 22 times with ROSAT in a period of seven years from 1991 
to 1998: 12 times with the PSPC 
\citep[Position Sensitive Proportional Counter;][]{pfef} and 10 times with the 
HRI \citep[High Resolution Imager;][]{david}. This frequent coverage is mainly
due to the monitoring of the supernova SN1993J \citep{zim}, 2.7' south of the nucleus
of M81, whose evolution was followed with roughly one observation every six 
months. The large field of view of the PSPC ($2^{\circ}$ diameter) and of the 
HRI (30' diameter)
ensures that the nucleus is included in every observation of SN1993J. 
The data have been processed using the IRAF(v. 2.11)/PROS (v. 2.5) 
software system \citep{tody,wor}.

The radius for source photon selection for the PSPC depends on the position of 
the source on the detector plane, because the radius of the Point Spread
Function (PSF) increases with the distance 
from the center of the field of view (FOV) \citep{boese}; for the on-axis 
observations we used a source radius of 3', which includes 95\% of the emission 
at all energies and an annular background region with inner and outer radii 3'
and 7' respectively. For the off-axis observations the source position and radius 
were evaluated with the {\sc galpipe} processing \citep{dami} applied to each 
observation and from the ROSAT/PSPC point spread function description
\citep{boese}: 
the radius is 5' and 6', for M81 nucleus off-axis positions of 16' and  
36' to 42', respectively. We have excluded any part of the
selected region obscured by the PSPC window supporting ribs.  
The background was extracted from an annular region with outer and inner radii 
12' and 6' respectively, after subtraction of the contribution of point sources. 
For convenience of reference, we designate the PSPC pointings P1-P12 in
time order.

The ten HRI observations pointed on M81 (H1-H10) have a total exposure 
time of 177 
ksec. For each observation we used the position of the detected 
sources\footnote{using the detection algorithm in the IRAF/PROS task 
{\sc xexamine}}, excluding the nuclear source M81 X-5, to check for possible
misalignements of the astrometric frames and to correct for small spacecraft 
errors; no further alignment was needed. The PSPC and HRI position 
determinations are consistent with each other and from now on we will use the 
more accurate HRI centroid coordinates: RA $9^h 57^m 54.3\pm 0.1^s$ and Dec 
$69^{\circ}03'46.4\pm 0.5"$ (J2000). For each HRI observation, source counts 
were extracted from a circular region with 1.3' radius, as found with the 
{\sc ldetect} IRAF/PROS algorithm (S. Dyson 1999, private communication). 

The 21 available ASCA observations of M81 (A1-A21) were also used to estimate 
the flux of
the nucleus (see Table~\ref{log}; for a detailed analysis of these data see
\citealp{ishi} and \citealp{iyo}). The ASCA/SIS data are less contaminated by the
SN1993J emission than the ASCA/GIS ones. We extracted ASCA/SIS spectra from a 
4' circular region centered 
on the apparent centroid of the M81~X-5 emission, excluding a 2' circular region
centered on the supernova (that lies 2.8' from M81~X-5). The background was
extracted from the portion of the SIS chip not contained in the above regions.

BeppoSAX observed M81 on June 4 1998. This observation was studied in detail by
\citet{pelle}. Here, we compare the SAX/LECS (Low Energy Concentrator Spectrometer) 
spectrum and ROSAT/PSPC spectra in order to check for spectral variability. 
For the BeppoSAX data, we used the standard
source and background spectra provided by the Narrow Field Instruments public 
archive\footnote{Available at http://www.asdc.asi.it/bepposax/}, extracted from 
a 6' radius region centered on the source centroid position.

\cha/ACIS has observed M81 twice (\citealp{ho4}, \citealp{swa}). We used the 
observation with the longest exposure time (50 ksec, Table~\ref{log}) . The 
high flux of the nucleus produces $>$ 80\% of pile-up fraction, making the 
direct study of the properties of the nucleus 
unreliable. We therefore used the read-out trailed image of the nucleus
to extract a spectrum unaffected by 
pile-up, using a
narrow box region (5'') running along the read-out direction of the chip 
containing the nuclear trailed image and excluding the direct image. 
The background was extracted from two similar regions adjacent to the source 
region, each 5'' wide. In the following, this spectrum will be referred to as 
\cha-T.  The sub-arcsecond \cha PSF allows the study of the circum-nuclear
region, and in particular of its contribution to the ROSAT and BeppoSAX spectra. To this aim
we extracted separately the spectra of the point-like sources (spectrum {\bf a})
and of the unresolved emission (spectrum {\bf b}), using the same extraction radius as for the on-axis ROSAT 
spectra (3'). The coordinates of the point-like sources are from \citet{swa}.
We excluded the point-like nuclear source (a circular region with a 10'' radius)
and the read-out trailed image (from a strip 5'' wide running along through the
chip read-out direction). The background is extracted from the remaining portion
of the chip. 

\clearpage
\begin{deluxetable}{rrrrr|rrrrr}
\tabletypesize{\footnotesize}
\tablecaption{Log of X-ray observations of M81.\label{tbl-1}}
\tablewidth{0pt}
\tablehead{
\colhead{Ref.\tablenotemark{a}}     &
\colhead{Obs. id.}   & \colhead{Live Time}           &
\colhead{Start Date} & \colhead{Angle\tablenotemark{b}}&
\colhead{Ref.\tablenotemark{a}}     &
\colhead{Obs. id.}   & \colhead{Live Time}           &
\colhead{Start Date} & \colhead{Angle\tablenotemark{b}}}
\startdata
 1P  &rp600101a00&    9296&25/03/91&  0.3'& 1A  & 15000000  &  6480& 05/04/93& 2.9'\\
 2P  &rp600110a00&   12717&27/03/91& 36.9'& 2A  & 15000120  & 30144& 05/04/93& 2.9'\\
 3P  &rp600052n00&    6588&18/04/91& 41.7'& 3A  & 15000010  &  6622& 07/04/93& 10.1'\\
 4P  &rp600110a01&   12238&15/10/91& 36.9'& 4A  & 15000130  & 35680& 07/04/93& 6.7'\\
 5P  &rp600101a01&   11085&16/10/91&  0.3'& 5A  & 15000030  &154847& 16/04/93& 6.6'\\
 6P  &rp600382n00&   27120&29/09/92&  0.3'& 6A  & 15000020  & 29440& 25/04/93& 6.5'\\
 7P  &rp180015n00&   17938&03/04/93&  0.3'& 7A  & 15000040  & 28576& 01/05/93& 6.6'\\
 8P  &rp180015a01&    8731&04/05/93&  2.8'& 8A  & 15000050  & 85376& 18/05/93& 5.9'\\
 9P  &wp600576n00&   16412&29/09/93& 16.0'& 9A  & 10018000  & 45152& 24/10/93& 4.0'\\
10P  &rp180035n00&   17800&01/11/93&  0.3'&10A  & 51005000  & 80016& 01/04/94& 6.6'\\
11P  &rp180035a01&    4234&07/11/93&  0.3'&11A  & 52009000  & 98206& 21/10/94& 7.5'\\
12P  &rp180050n00&    1849&31/03/94&  0.3'&12A  & 53008000  & 24128& 01/04/95& 7.2'\\
 1H  &rh600247n00&   26320&23/10/92&  0.3'&13A  & 53008010  &101434& 24/10/95& 7.6'\\
 2H  &rh600247a01&   21071&17/04/93&  0.3'&14A  & 54008000  &103616& 16/04/96& 7.0'\\
 3H  &rh600739n00&   19902&19/10/94&  0.3'&15A  & 54008010  & 39712& 27/10/96& 7.7'\\
 4H  &rh600740n00&   18984&13/04/95&  0.3'&16A  & 55018000  &100640& 08/05/97& 6.5'\\
 5H  &rh600881n00&   14826&12/10/95&  0.3'&17A  & 56020000  &102672& 10/04/98& 6.9'\\
 6H  &rh600882n00&   18328&15/04/96&  0.3'&18A  & 56020010  & 98224& 20/10/98& 7.6'\\
 7H  &rh600882a01&    5091&27/10/96&  0.3'&19A  & 57048000  & 79794& 06/04/99& 7.4'\\
 8H  &rh601001n00&   19231&29/03/97&  0.3'&20A  & 57006000  & 97104& 10/04/99& 6.7'\\
 9H  &rh601002n00&   12590&30/09/97&  0.3'&21A  & 57006010  & 99808& 20/10/99& 7.6'\\
10H  &rh601095n00&   12590&25/03/98&  0.3'&1S	& 40732001  &100000& 04/06/98& 0.0'\\
     &           &        &        &      &Ch   & 735       & 50570& 07/05/00& 2.8'\\ \tableline
\enddata
\tablenotetext{a}{Reference number for each observation. Instrumests are coded as: 
P=ROSAT/PSPC; H=ROSAT/HRI; A=ASCA; S=SAX; Ch=\cha}
\tablenotetext{b}{Nominal off-axis angle of the optical center of M81 on the 
detector}
\label{log} 
\end{deluxetable}
\clearpage

\section{ROSAT TEMPORAL ANALYSIS}
\label{time}
\begin{figure}
\centerline{\psfig{figure=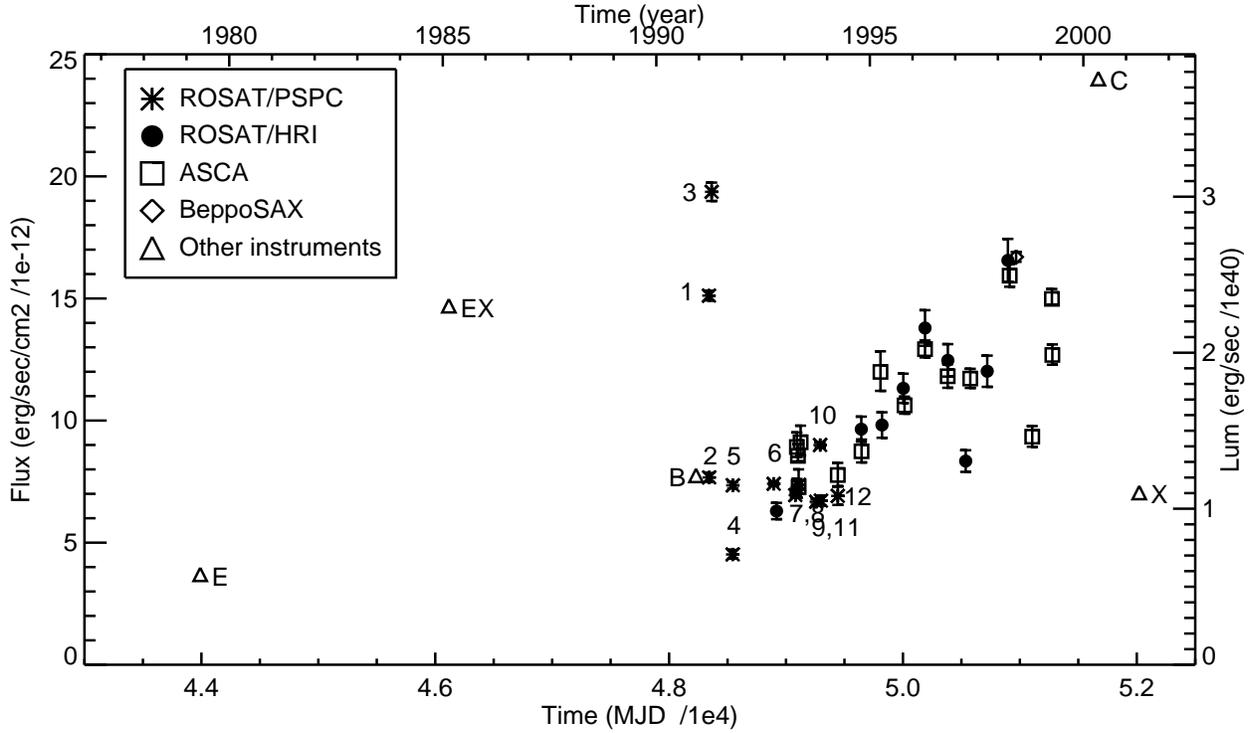,width=18cm}}
\caption{Observed flux (not corrected for cold absorption) in the 0.5-2.4 keV 
band (left Y axis). The details on
models used for flux calculation and spectral analysis are given in 
Sections~\ref{time} and \ref{spec}. The luminosity 
(right Y axis) is calculated assuming a distance of 3.6 Mpc \citep{fre}. 
The points from other instruments are derived from the literature and are marked
as follows:
E~=~{\it Einstein}/IPC \citep{fab88}; EX~=~EXOSAT and B~=~BBXRT \citep{petre};
C~=~\cha/ACIS \citep{swa}; X~=~XMM \citep{page}
The numbers near the symbols of the twelve PSPC observations show the time 
sequence of the relevant data points and clarify the rapid and large transients 
in the first part of the light curve.}
\label{flux}
\end{figure}

The ROSAT data form a long series of uniform observations, suitable for temporal
analysis. Instead, the wide beam of ASCA and BeppoSAX data includes
much galaxian emission, and we do not include them in the following analysis,
except for the long-term light-curve. In Figure~\ref{flux} we show the 
1991-1999 long term light curve of the nucleus of M81
in the 0.5-2.4 keV band. The PSPC fluxes were calculated assuming a power-law
with $\Gamma=1.79$ plus a thermal component with $kT\sim 0.5$ keV, obtained as
final result of the spectral analysis (see Section~\ref{spec}). The same model 
was used to calculate the fluxes from the ROSAT/HRI count rates, through the 
{\sc PIMMS} tool\footnote{Portable Interactive Multi-Mission Simulator, 
at http://asc.harvard.edu/toolkit/pimms.jsp}. 
ASCA/SIS fluxes were evaluated using a 
power-law model, since the soft thermal component is negligible in the ASCA
spectral band; the BeppoSAX flux was extrapolated from the value 
reported in \citet{pelle} for the 0.1-2.0 keV band using their best-fit model. 
We note that the light curve in Figure~\ref{flux} has been derived 
using data from four different detectors, and therefore one may expect some 
cross-calibration problems. In our case this is not a problem, since
data taken with different instruments  very close in time give consistent
results. This happens in most HRI/ASCA pairs of data (see, for example, the
groups of observations at $5.00\times10^4 d$, $5.02\times10^4 d$, 
$5.04\times10^4 d$); see also the group of observation at $5.09\times10^4 d$ 
which consists of HRI, ASCA and BeppoSAX points. 

We observe a factor of 2-4 flux difference between the
two high count-rate PSPC observations 1P and 3P and all the other PSPC 
pointings (all of which yield a similar, lower, count rates). 
We can confidently exclude an unlikely variability in the instrumental
calibration, because no flux enhancement is seen in any of the sources in the
field (see, e.g., X-9 \citealp{lapa} and other fainter sources \citep{imm}.
We can also rule out transient sources near the nucleus, because there is no 
significant spectral variation between 1P and 3P and the immediately following 
pointings (Section~\ref{spec}). The above considerations 
suggest that this variation should be ascribed entirely to the nuclear source.

To search for short term variations, we examined the light-curves from 
individual ROSAT observations. Figure~\ref{ltc} presents these light-curves in 
order of their observation. Each bin corresponds to one GTI (Good Time Interval
\footnote{Most targets are occulted by Earth for part of the orbit and the
observations are broken into smaller 
segments which may be interleaved with observations of other targets. Hence, the
resulting data stream for a particular target will in general show large data 
gaps, with effective observing intervals lasting a few hundreds of seconds;
each of these continuous observation is a ``Good Time Interval''.})
and the time is given in days starting from the ROSAT launch (June 1990).

The data show variability on three different timescales:\\
\ibu Slow variability: a slow regular ascending trend can be observed,
starting on November 1993 and ending on April 1996, with a difference of a 
factor of $\sim2$ 
in the flux between the first and the last point (Figure~\ref{flux}). This 
trend was reported by \citet{pelle} and \citet{iyo} for the 2.0-10.0 keV band 
from BeppoSAX and ASCA data.\\
\ibu Medium variability: both before and after the period of regularly
increasing flux, there are two periods of irregular variability 
(Figure~\ref{flux}). This is more
marked before November 1993, where two very bright flare-like episodes are 
seen, with variations up to 4 times in flux. It is weaker (but nevertheless 
extremely significant) after April 1996. This kind of variability was not 
previously reported.\\
\ibu Fast variability: the lightcurves of many observations (Figure~\ref{ltc}) 
show variability of up to 30\% on timescales of the order of one day (e.g. 
observations 9P, 4H, 6H, 9H), as found with BeppoSAX \citep{pelle}, or even a 
few hours (e.g. observations 5P, 9P, 10H), confirming the early EXOSAT report 
\citep{barr}.

A Kolmogorov-Smirnov variability test \citep[KS, see, e.g.,][]{cono} applied to
the unbinned data and a $\chi^2$ test applied to the light curves binned into
GTI show that 11 out of 22 observations are variable in both tests with more 
than 99\% probability of rejecting the hypothesis of constant rate 
(Table~\ref{tflux}).
Despite the abrupt drop in flux by a factor of $\sim3$ in one day between 
observations 1P and 2P and the factor of $\sim2$ rise  
20 days later between 2P and 3P (Figures~\ref{flux} and \ref{ltc}), the 
individual light curves of these observations do not show any sharp change of 
the count rate. 

\begin{figure}
\centerline{\psfig{figure=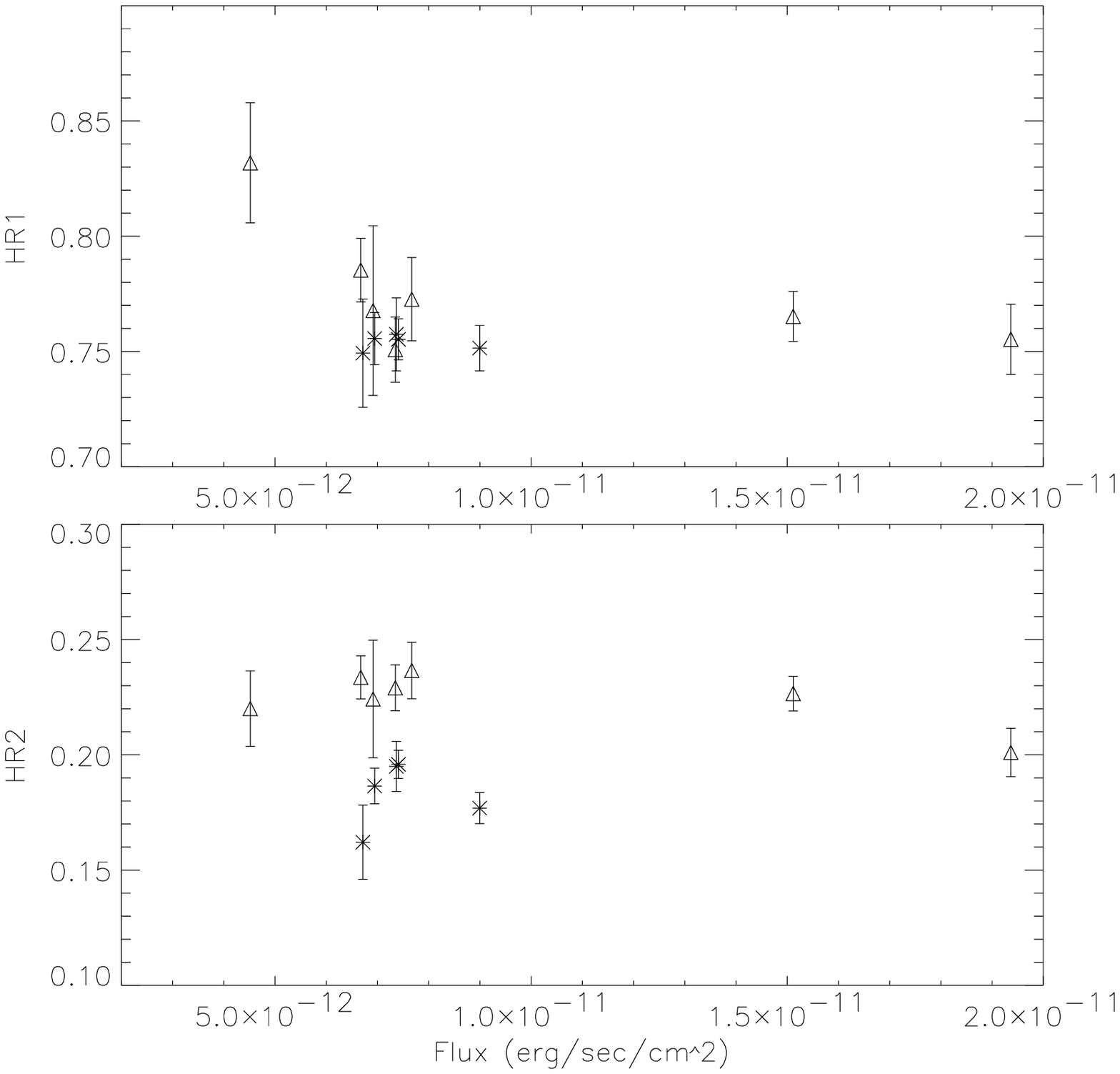,width=15cm}}
\caption{Hardness ratios (as defined in Table~\ref{tflux}) vs. count rate.
Triangles and stars indicate respectively the observations with and without the
strong absorption feature (see Sections~\ref{indspec} for a
discussion)}
\label{hr}
\end{figure}

We investigated the spectral variability in the PSPC observations by 
calculating a hardness ratio of each observation in two pairs of bands
(Table~\ref{tflux}). The 
first, HR1: 0.11-0.42 keV/ 0.52-2.02 keV, covers the whole spectral range of the
PSPC and can give information on the absorbing column. The second HR2: 
0.52-0.91 keV/ 0.92-2.02 keV, covers only the hardest part of the spectrum and
is more sensitive to spectral index variations. We performed a $\chi^2$ test
against the ipothesis of constant distribution centered on the average value: 
HR1 does not show any variability ($\chi^2/\nu=13.6/11$, where $\nu$ is the 
number of degrees of freedom) while HR2 shows evidence of variability 
($\chi^2/\nu=65/11$). We found no evidence of 
correlation of the hardness ratio with the total flux (Figure~\ref{hr}).
The spectral analysis revealed that the spectrum is indeed variable, 
as discussed in Section~\ref{spec}.

%chisq results for hr vs. constant mean: 
%HR1... xsq=      23.2883  red xsq=     0.226100  prob=      1.00000
%HR2... xsq=      102.172  red xsq=     0.991964  prob=     0.504515  
%source dimensions from variability: 1e-3pc
%source dimensions from image: <170pc
\clearpage
\begin{deluxetable}{crrrrrr}
\tabletypesize{\footnotesize}
\tablecaption{Properties of individual ROSAT observation of M81.}
\tablewidth{0pt}
\tablehead{
\colhead{Ref.}  &
\colhead{$\Gamma$\tablenotemark{a}} & \colhead{Flux\tablenotemark{b}}&   
\colhead{P$_{KS}$\tablenotemark{c}}&\colhead{P$_{\chi^2}$\tablenotemark{d}}&
\colhead{HR1\tablenotemark{e}} & \colhead{HR2\tablenotemark{e}}}
\startdata
 1P &$2.20\pm0.08 $ &15.12 &0.21    & 0.26   &$0.765\pm0.011$&$0.227\pm0.007$\\
 2P &$1.92\pm0.15 $ &7.67  &$<0.01 $&$<0.01 $&$0.773\pm0.018$&$0.237\pm0.012$\\
 3P &$2.10\pm0.12 $ &19.37 &0.49    & 0.01   &$0.755\pm0.015$&$0.201\pm0.010$\\
 4P &$2.1\pm0.2 $   &4.51  &$<0.01 $&$<0.01 $&$0.832\pm0.026$&$0.220\pm0.016$\\
 5P &$2.22\pm0.10 $ &7.34  &0.16    &$<0.01 $&$0.751\pm0.014$&$0.229\pm0.010$\\
 6P &$2.30\pm0.07 $ &7.41  &$<0.01 $&$<0.01 $&$0.755\pm0.009$&$0.196\pm0.006$\\
 1H & \nodata       &6.29  &$<0.01 $&$<0.01 $& \nodata       & \nodata\\
 7P &$2.32\pm0.09 $ &6.94  &$<0.01 $&$<0.01 $&$0.756\pm0.011$&$0.186\pm0.008$\\
 2H & \nodata       &7.22  &0.53    & 0.71   & \nodata       & \nodata\\
 8P &$2.25\pm0.12 $ &7.37  &$<0.01 $&$<0.01 $&$0.757\pm0.016$&$0.195\pm0.011$\\
 9P &$2.18\pm0.10 $ &6.67  &$<0.01 $&$<0.01 $&$0.785\pm0.014$&$0.234\pm0.009$\\
10P &$2.41\pm0.06 $ &9.00  &0.55    & 0.50   &$0.751\pm0.010$&$0.177\pm0.007$\\
11P &$2.47\pm0.20 $ &6.71  &0.02    & 0.03   &$0.749\pm0.023$&$0.162\pm0.016$\\
12P &$2.2^{+0.4}_{-0.3}$&6.91  &0.06& 0.14   &$0.768\pm0.037$&$0.224\pm0.025$\\
 3H & \nodata       &9.64  &0.16    & 0.40   & \nodata       &\nodata\\
 4H & \nodata       &9.81  &$<0.01 $&$<0.01 $& \nodata       &\nodata\\
 5H & \nodata       &11.32 &$<0.01 $&$<0.01 $& \nodata       &\nodata\\
 6H & \nodata       &13.79 &$<0.01 $&$<0.01 $& \nodata       &\nodata\\
 7H & \nodata       &12.46 &$<0.01 $&$<0.01 $& \nodata       &\nodata\\
 8H & \nodata       &8.34  &0.02    & 0.03   & \nodata       &\nodata\\
 9H & \nodata       &12.02 &$<0.01 $&$<0.01 $& \nodata       &\nodata\\
10H & \nodata       &16.56 &$<0.01 $& 0.09   & \nodata       &\nodata\\ \tableline
\enddata
\tablenotetext{a}{Best fit power-law photon index}
\tablenotetext{b}{Average flux in units of $10^{-12}$ erg s$^{-1}$ cm$^{-2}$. The
model used to calculate fluxes are described in Section~\ref{time}.}
\tablenotetext{c}{Probability of a constant count rate within the observation 
according to a Kolmogorof-Smirnof test}
\tablenotetext{d}{Probability of a constant count rate equal to the average
count rate according to a $\chi^2$ test}
\tablenotetext{e}{Average PSPC hardness ratios defined as
$HR=\frac{H-S}{H+S}$, where $H_{HR1}=C_{0.52-2.02}$, 
$S_{HR1}=C_{0.11-0.42}$, $H_{HR2}=C_{0.92-2.02}$, 
$S_{HR2}=C_{0.52-0.91}$
%$HR1=\frac{C_{0.11-0.42}-C_{0.52-2.02}}{C_{0.11-0.42}+C_{0.52-2.02}}$ and
%$HR2=\frac{C_{0.52-0.91}-C_{0.92-2.02}}{C_{0.52-0.91}+C_{0.92-2.02}}$ where $C$
are the photon counts in the subscripted energy bands (in keV).}
\label{tflux} 
\end{deluxetable}
\clearpage

\begin{figure}
\centerline{\psfig{figure=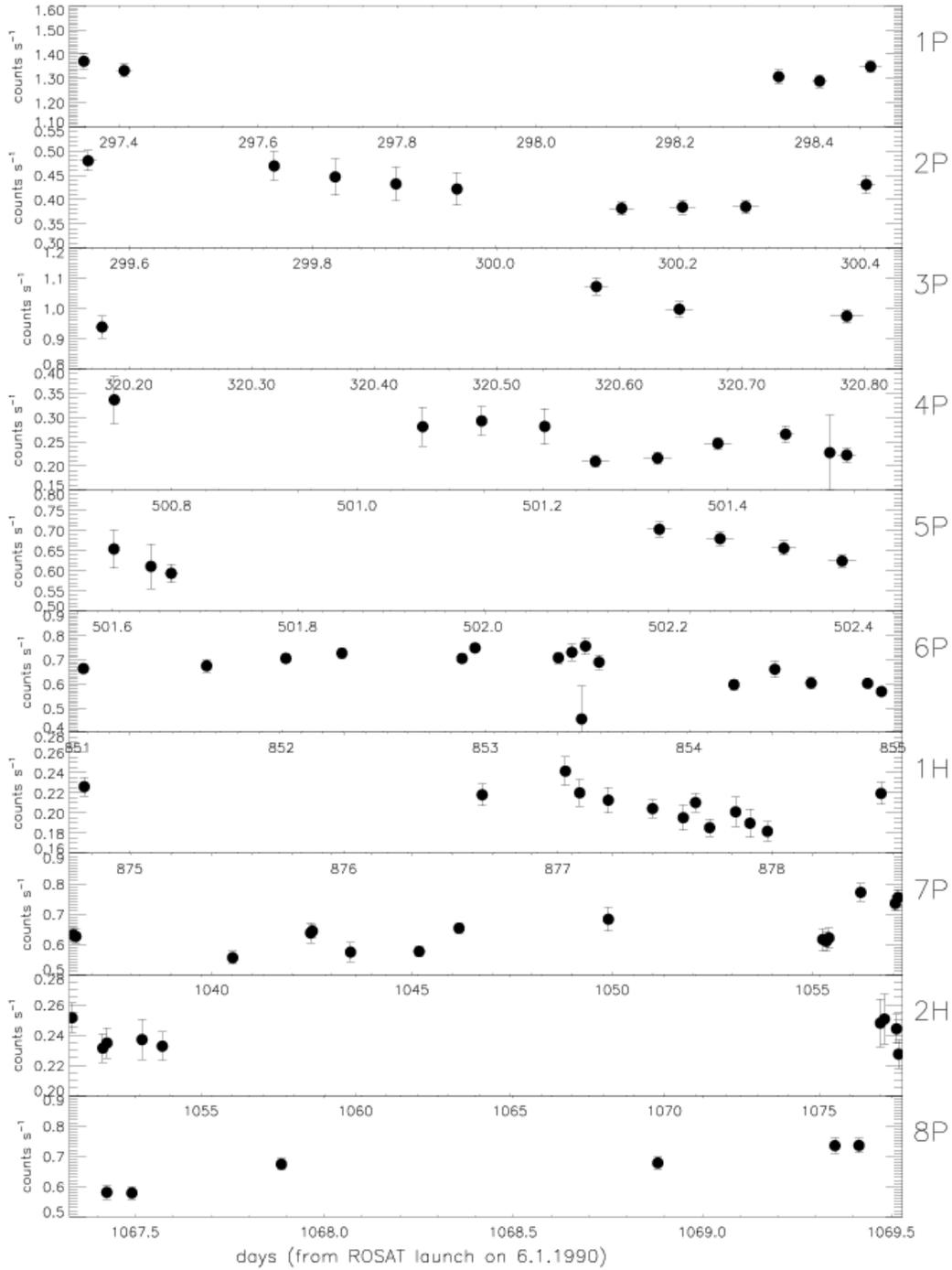,width=15cm,angle=90}}
\caption{Light curves of PSPC and HRI observations (identified with a P and a H
on the right of the graph
respectively). Time is in days since the satellite launch (June 1990); the 
number on the right of each curve identifies the observations sequence as in 
Table~\ref{log}.}
\label{ltc}
\end{figure}
\begin{figure}
\figurenum{\ref{ltc}}
\centerline{\psfig{figure=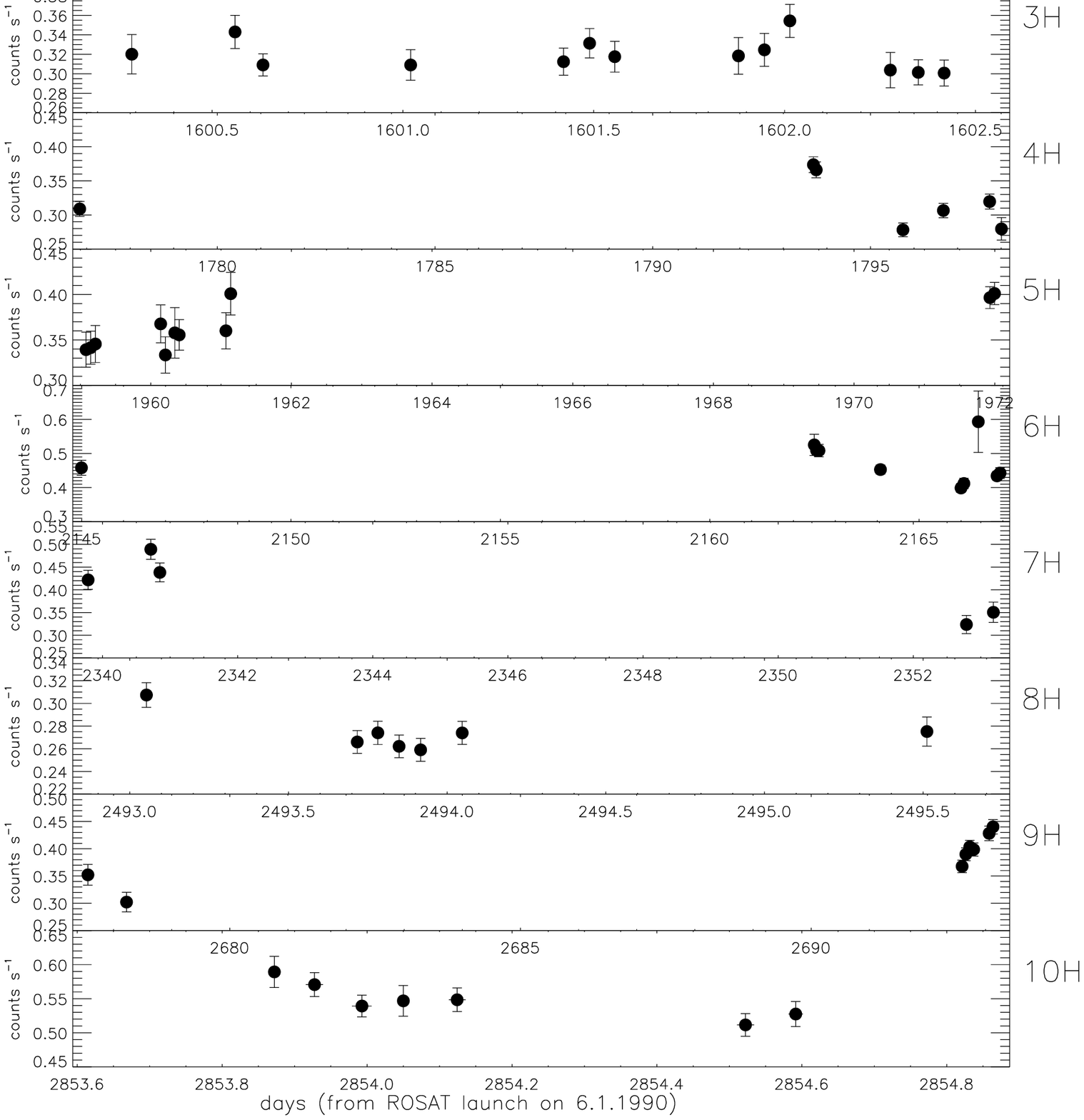,width=15cm,angle=90}}
\caption{\it Continued}
\label{ltc2}
\end{figure}

\clearpage
\section{SPECTRAL ANALYSIS}
\label{spec}
The spectra were analysed using the software 
XSPEC{\sc v 10.0/11.1} \citep{arn}.
\subsection{The galactic emission}
\label{chaspec}
The large beam spectra from BeppoSAX and ROSAT include a significant 
contribution from the circumnuclear region of the galaxy, within the 3' 
extraction radius around the
nucleus. The high resolution \cha observation allows us to measure the 
galaxy emission (both diffuse and from individual point sources) included in 
the wider beam ROSAT and BeppoSAX spectra. As described in Section~\ref{data}, 
we used the same extraction radius as for the ROSAT data (3'), but excluded the 
region affected by the nuclear point source and the strip containing the 
trailed image of the nucleus. We
then extracted two spectra: {\bf a)} the individual point source spectrum, 
obtained summing the spectra of the bright sources falling within the 3' 
circle, as listed in \citet{swa}, {\bf b)} the diffuse spectrum, i.e. the 
emission within the same region excluding the point sources. These have roughly 
equal fluxes. We then formed  the total spectrum {\bf c)}, 
summing the diffuse and the point-source spectra.
We fitted simultaneously the spectra {\bf a)}, {\bf b)} and {\bf c)}, imposing
the best fit model of spectrum  {\bf c)} to be the sum of the best fitting model
of spectra {\bf a)} and {\bf b)}.
%We analyzed spectra {\bf a)} and {\bf b)} separately, and then merged the the
%models of {\bf a)} and {\bf b)} to fit spectrum {\bf c)}. More in detail, we
%used the results of both {\bf a)} and {\bf b)} as initial values for the fitting
%and we fitted simultaneously {\bf a)}, {\bf b)} and {\bf c)} with the constrain
%that the model fitting {\bf c)}has to be the sum of those fitting {\bf a)} and 
%{\bf b)}. 
The best fit parameters are in Table~\ref{fit}.
We find that spectrum {\bf a)} can be described with power-law with
$\Gamma\sim1.7$ plus a multicolor black body (MCBB, model {\sc diskbb} in {\sc
xspec}) at $\sim0.08$ keV and a gaussian at $\sim0.8$ keV. To describe spectrum 
{\bf b)} we need a Raymond-Smith spectrum at $\sim 0.25$ keV and, again, a 
power-law with $\Gamma\sim1.7$. Spectra {\bf a)} and {\bf b} are not consistent 
with each other. However, in both cases the N{\sc h} is
consistent with the Galactic line of sight value and the two power-law slopes
are consistent with each other. In merging the models to fit spectrum {\bf c} we
used only a single power-law component. The [0.5-2.4] keV flux corrected for
absorption by the best fit N{\sc h} is 
$8.1\times10^{-13}$ \flux~for the integrated point source 
emission and $6.9\times10^{-13}$ \flux~ for the diffuse emission, for a total of
$1.5\times10^{-12}$ \flux~. 

\subsection{The nuclear spectra}
\label{indspec}
\begin{figure}
\centerline{\psfig{figure=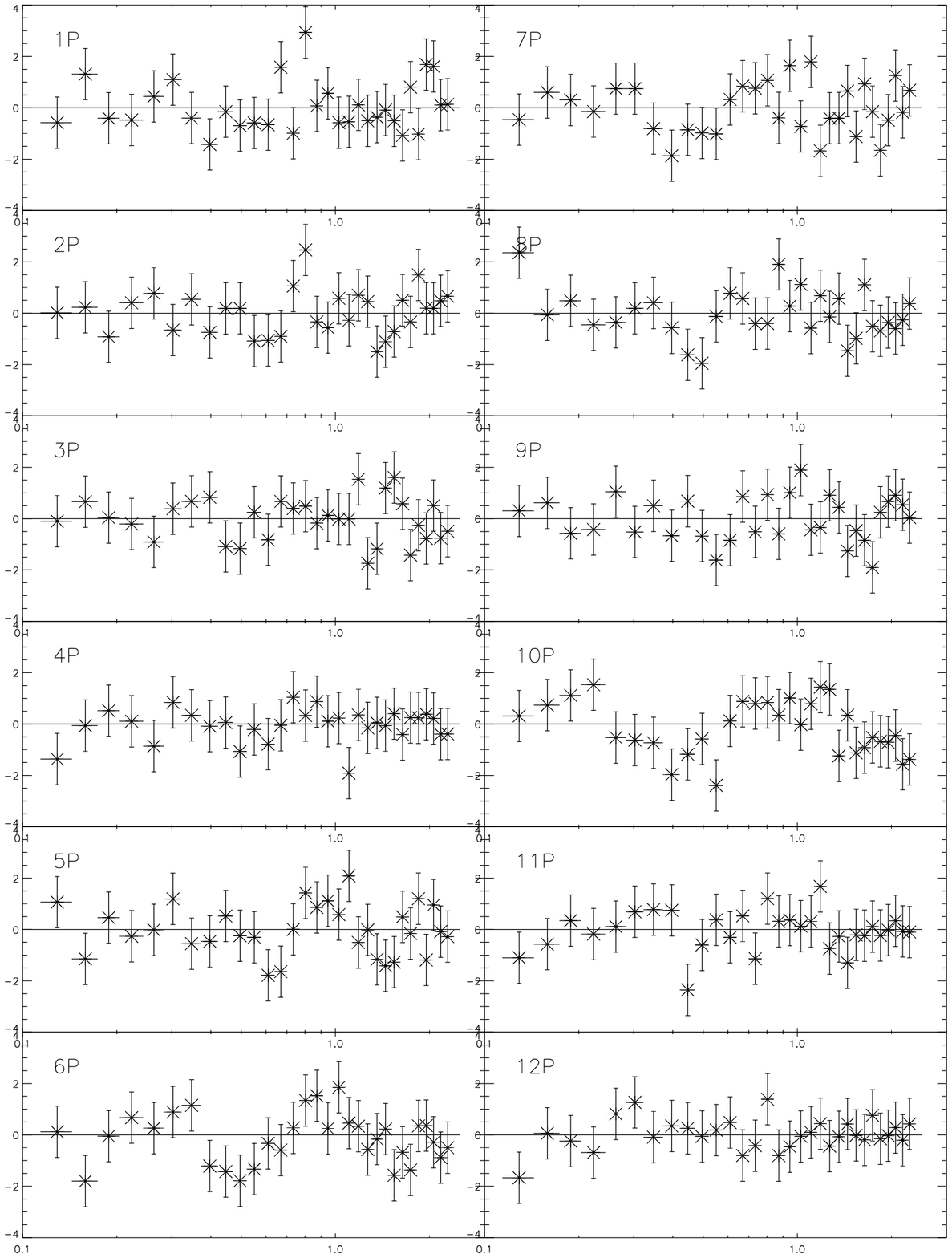,width=15cm,angle=0}}
\caption{Residuals in units of $\chi$ for all the PSPC spectra best-fitted with 
a power law model}
\label{resid}
\end{figure}

We first analysed the ROSAT spectra, extracted with the IRAF routine {\sc qpspec} and 
processed with {\sc proscon}, using the response matrices publicly available
through the HEASARC/ROSAT webpage\footnote{http://heasarc.gsfc.nasa.gov/docs/rosat/rosgof.html}. 
As a preliminary step, we fitted the spectra of each of the 12 observations 
with a simple absorbed power law model, in order to search for spectral 
variability. 
A comparison of residuals of each observation to the relevant best fit model
(Figure~\ref{resid}) shows that some observations (6P, 7P, 8P, 10P, hereafter ``dip'' observations) show a large dip in the 
residuals between 0.2 and 0.7 keV, while the remaining seven observations 
(hereafter ``flat'') are well described by a power-law. 

\begin{figure}
\centerline{\psfig{figure=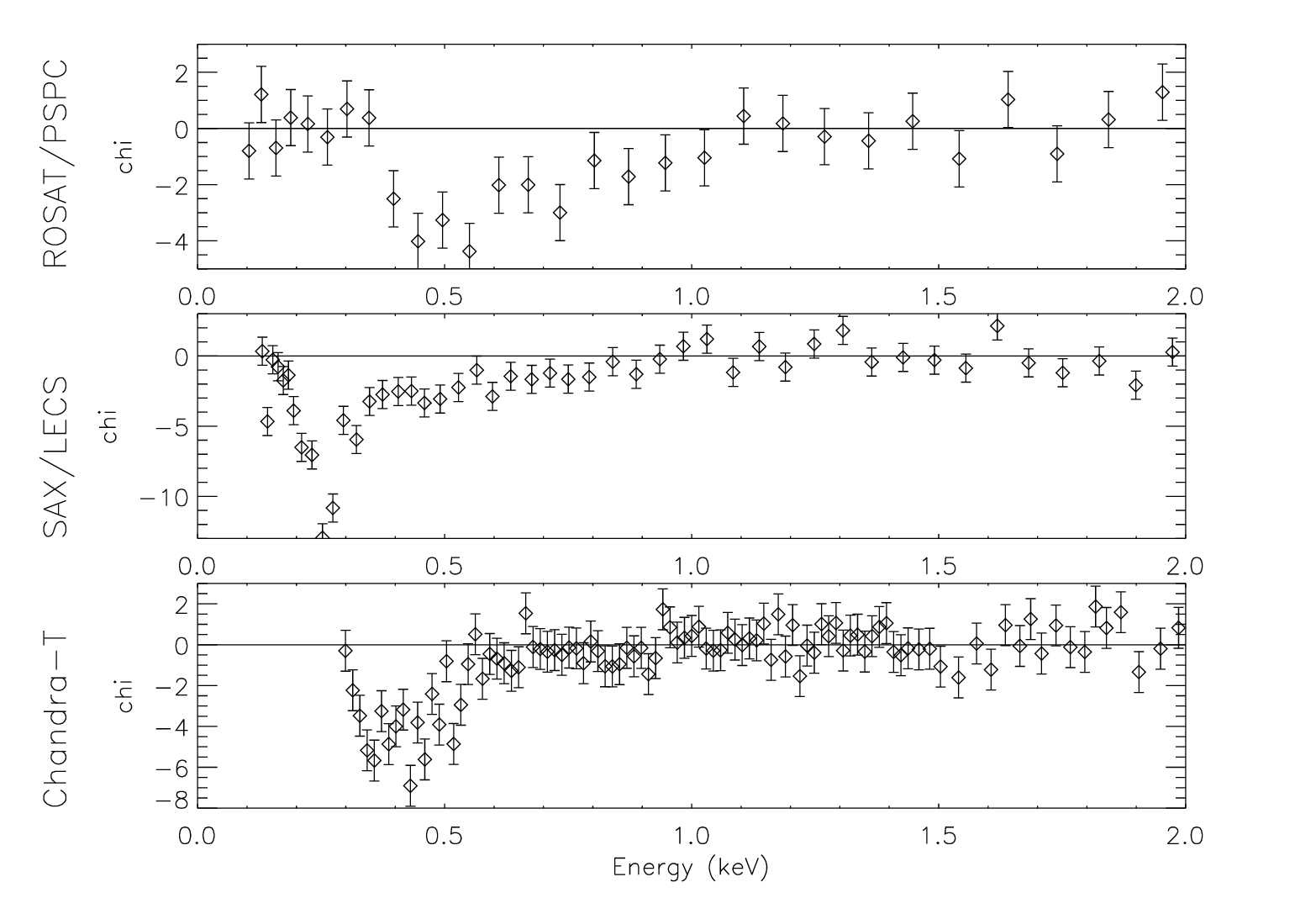,width=15cm,angle=0}}
\caption{Spectra of the nucleus of M81: residuals with respect to a power-law
model. Form top to bottom: ROSAT/PSPC  
spectrum (obtained from the sum of the on-axis observations), SAX/LEX spectrum, 
\cha-T spectrum. These residuals have been obtained
fitting with a power law energy range not affected by the absorption feature:
0.1-0.3 keV and 1.0-2.4 keV for the PSPC and $>1.0$ keV for LECS and \cha-T}
\label{phspec}
\end{figure}

We improved 
the statistics of the spectral fit by summing the spectra by type. We 
then fitted the two resulting spectra  
with a simple power-law. The results show that in the summed {\it dip} spectrum 
the absorption feature is clearly visible between 0.3 and 0.5 keV 
(Figure~\ref{phspec}); however, in the summed {\it flat} spectrum there
is also a broad, weak absorption-like structure, at the same energy, suggesting
that the feature is in fact present in all the observations, albeit with
varying intensity. We can confidently rule
out the hypothesis that the feature is an instrumental effect, as there is no
hint of it in the spectra of other sources of the same field (see e.g. X-9 in
\citealp{lapa}). In order to minimize the contribution from other bright sources
in the vicinity of the nucleus, and because the effective area is poorly
calibrated for off-axis PSPC sources, the subsequent analysis was carried on a
spectrum selected from the sum of the eight observations where the nucleus was 
on-axis.

We  also checked for  the presence of the absorption feature in observations by 
other instrument. ASCA has too little response below 0.5 keV. We can however
use SAX/LECS and {\it Chandra}-T (read-out trail) spectra (see 
Section~\ref{data} for details on 
the extraction of these spectra). We decided to fit the whole energy range
(0.1-4.0 for LECS, 0.3-8.0 for ACIS, compared with 0.1-2.4 for PSPC), in order
to have a better estimate of the continuum. We fitted each of these spectra 
with a simple power-law model (Table~\ref{fit2}). An absorption feature at 
$\sim0.3$ keV is seen in these spectra
as well. This is made more evident in Figure~\ref{phspec}, where we plot the
residuals obtained by fitting with a power law only the energy range above 1.0
keV, that shows also how the feature is at slightly different energy in the 
three spectra.

We then made a more careful analysis (see Table~\ref{fit2}) of the ROSAT/PSPC 
and SAX/LECS data by adding a fixed component to model the extended galaxy 
contribution, set to the best fit model derived from the \cha data of the 3' 
radius extraction region around the
nucleus (see Section~\ref{chaspec}) in addition to the variable components. 
A power-law plus this extended component
gives a good fit to energies higher than 1 keV, but still overpredicts the
emission between 0.3 and 0.6 keV. To investigate the nature of this ``dip'' we 
tested four models
in which a single component was added to the power-law: an optically thin 
Raymond-Smith plasma, a multi-color black body accretion disk (MCBB),
an absorption edge and a gaussian line in absorption (Table~\ref{fit2}). 
In all cases the 
improvement over a simple power-law model is highly significant, with F-test
probabilities lower than 1\%. The BeppoSAX and ROSAT data give similar results
for both the MCBB and the Raymond-Smith components, with a better $\chi^2$ 
for the disk model. In both
cases, the edge model has a $\chi^2$ comparable with that of the disk model. 
The energy of the edge
found for BeppoSAX data ($0.22^{+0.03}_{-0.06}$ keV) is lower than for ROSAT data 
($0.41^{+0.08}_{-0.13}$ keV). In the
\cha data we found that the fits require a lower temperature Raymond or MCBB
component, while the edge energy is $0.29\pm 0.02$ keV. The gaussian model
significantly improves the fit of the \cha data ($\chi^2/\nu=104/127$) while
for the PSPC data it is equivalent to the other models and for BeppoSAX it is
slightly worse. The best fit energies of the gaussian absorption line are 
consistent with each other in the three instruments.

The power-law index in the ROSAT data is steeper than in the
SAX data by $\Delta\Gamma\sim 0.4$ to 1.0, depending on the spectral model. 
An offset of $\Delta\Gamma\sim 0.4$ has been observed in other objects
too (e.g. NGC5548 \citealp{iwa}), and is thought to be a 
systematic calibration offset. In this case the difference is sometimes larger
than this offset ( in NGC5548), suggesting that at least part of the steepening
is real, and not an instrumental effect.\\
The luminosity of the power-law source in the 0.5-2.4 keV range is 
$2.3\times 10^{40}$erg/s (corresponding to $1.6\times 10^{-11}$ \flux)
for BeppoSAX and $1.4\times 10^{40}$ erg/s ($1.0\times 10^{-11}$ \flux) for the 
PSPC data used in the spectral analysis. Since the contribution of 
circumnuclear emission from the
\cha analysis was included in the fit models for the ROSAT/PSPC and SAX/LECS 
data, these luminosities are truly representative of the nuclear source. The 
{\it Chandra}-T spectrum cannot be used directly to 
obtain a normalization, and thus a flux. We use here the estimate of
\citet{swa}, see Figure~\ref{flux}, which give $L_X=3.8\times10^{40}$ erg/s in
the 0.5-2.4 keV range (i.e. $2.6\times 10^{-11}$ \flux).

In summary, our analysis revealed that the nuclear spectrum is consistent with
a power-law plus a soft component that can be modelled either with an absorption
feature at $0.3\sim0.4$ keV or with thermal disk emission (MCBB model) at 
$0.2\sim0.3$ keV. The spectrum is absorbed in
excess of the Galactic N{\sc h}, while the spectrum of the circumnuclear region
(only from \cha/ACIS data) does not show any excess absorption. The
circumnuclear emission is modelled as two components, resolved point sources 
(described with a power-law plus a MCBB, with $L_X=1.2\times10^{39}$ erg/s) and 
diffuse emission (described with a power-law plus a Raymond-Smith model, with 
$L_X=1.0\times10^{39}$ erg/s), that contribute for a total of 
$L_X=2.2\times10^{39}$ erg/s (i.e. $1.5\times 10^{-12}$ \flux).
\clearpage

\begin{figure}
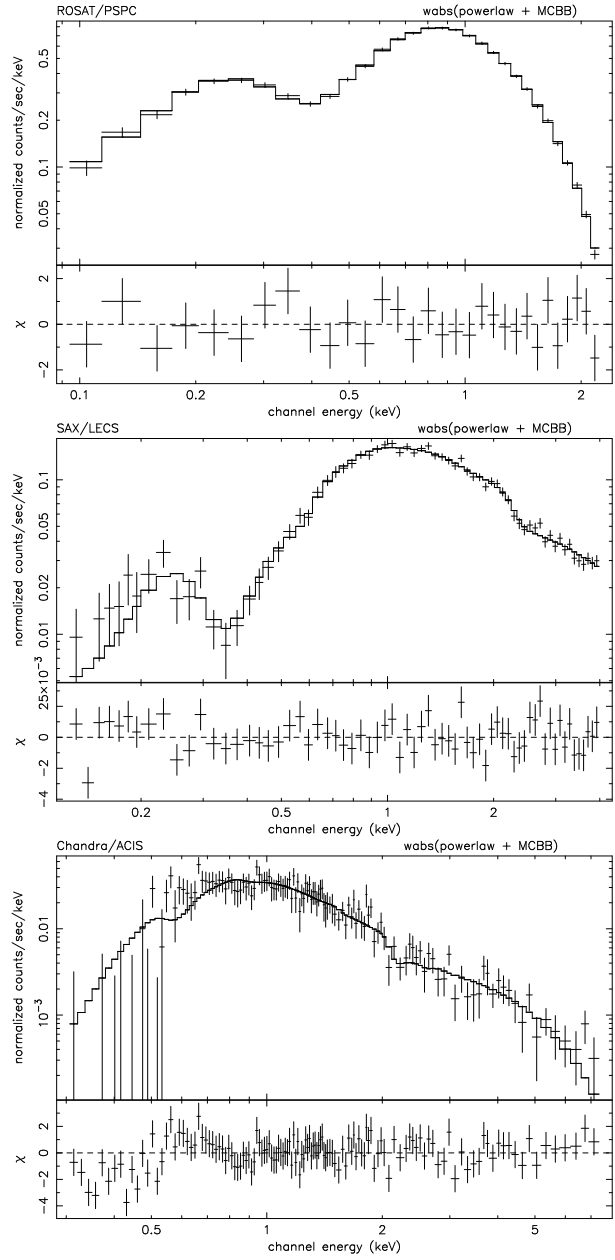

\centerline{\psfig{figure=f6a.ps,width=8cm,angle=270}}
\centerline{\psfig{figure=f6b.ps,width=8cm,angle=270}}
\centerline{\psfig{figure=f6c.ps,width=8cm,angle=270}}
\caption{Spectra, best fit model and residuals of the nucleus of M81 for the 
MCBB. {\bf Top}: PSPC; {\bf Middle}: SAX/LECS;  {\bf Bottom}: \cha-T}
\label{specps1}
\end{figure}

\begin{figure}
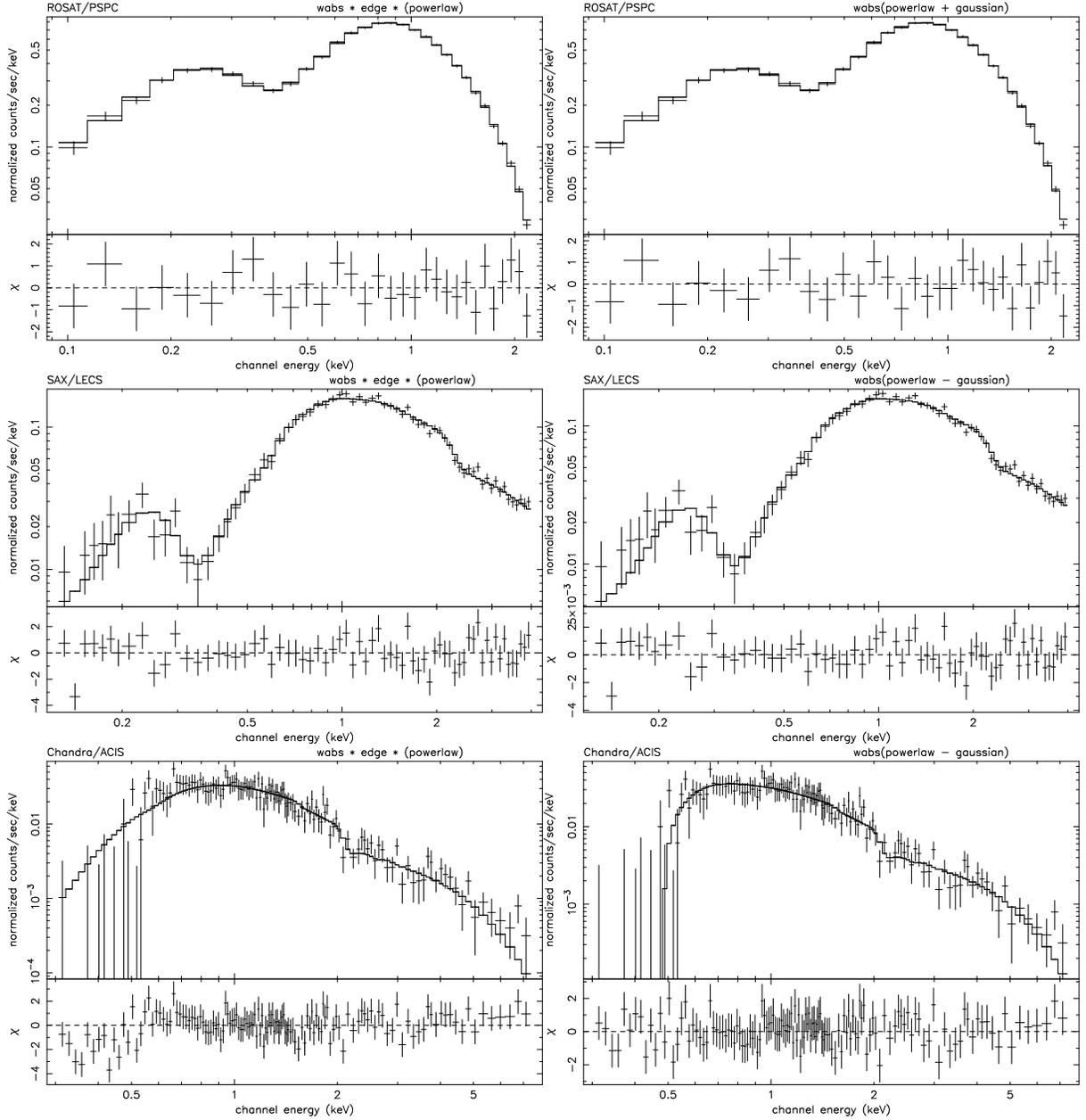

\centerline{\psfig{figure=f7a.ps,width=8cm,angle=270}\psfig{figure=f7b.ps,width=8cm,angle=270}}
\centerline{\psfig{figure=f7c.ps,width=8cm,angle=270}\psfig{figure=f7d.ps,width=8cm,angle=270}}
\centerline{\psfig{figure=f7e.ps,width=8cm,angle=270}\psfig{figure=f7f.ps,width=8cm,angle=270}}
\caption{Spectra, best fit model and residuals of the nucleus of M81 for the 
edge model (left column) and the gaussian model (right column). {\bf Top}: PSPC;
{\bf Middle}: SAX/LECS; 
{\bf Bottom}: \cha-T}
\label{specps}
\end{figure}

\clearpage
\begin{deluxetable}{lrrr}
\tabletypesize{\footnotesize}
\tablecaption{BEST FIT MODEL TO THE CIRCUMNUCLEAR GALACTIC EMISSION FROM \cha}
\tablewidth{0pt}
\tablehead{
\colhead{Component}&\colhead{Parameters}&\colhead{Point sources}& \colhead{Diffuse emission}\\
 & &\colhead{Spectrum {\bf a}}& \colhead{Spectrum {\bf b}}}
\startdata
          &$N_H$ ($10^{20}$ cm$^{-2}$)	&$4.1^{+0.4}$          &$4.1^{+0.4}$\\[0.2cm]
Power-law &$\Gamma$			&$1.74^{+0.05}_{-0.04}$&$1.74^{+0.05}_{-0.04}$\\[0.2cm]
          &Flux (\flux)                 &$6.5\times10^{-13}$   &$4.6\times10^{-13}$\\[0.2cm]
MCBB      & kT  (keV)                   &$0.083^{+0.003}_{-0.006}$&\nodata \\[0.2cm]
          &Flux (\flux)                 &$8.9\times10^{-14}$    &\nodata \\[0.2cm]
Gaussian  & E (keV)                     &$0.86^{+0.03}_{-0.04}$&\nodata \\[0.2cm]
          &$\sigma$ (keV)               &$0.15\pm0.03$         &\nodata \\[0.2cm]
          &Flux (\flux)                 &$7.0\times10^{-14}$    &\nodata \\[0.2cm]
Raymond-Smith&	   kT  (keV)            &\nodata               &$0.273\pm0.011$\\[0.2cm]
          &Flux (\flux)                 &\nodata               &$2.3\times10^{-13}$\\[0.2cm]
	  &$\chi^2/\nu$ 		&256/197               &169/126 \\[0.2cm]
\enddata
\tablecomments{The N{\sc h} is constrained to be not less than the Galactic 
line-of-sight value ($4.1\times10^{20}$ cm$^{-2}$). All fluxes are calculated 
in the 0.5-2.5 keV band and corrected for best fit N{\sc h}}
\label{fit} 
\end{deluxetable}
\begin{deluxetable}{lrrrr}
\tabletypesize{\scriptsize}
\tablecaption{BEST FIT MODELS FOR THE NUCLEAR EMISSION}
\tablewidth{0pt}
\tablehead{
\colhead{Model}&\colhead{Parameters}&\colhead{PSPC}& \colhead{SAX}& \colhead{\cha}}
\startdata
Power-law &$N_H$ ($10^{20}$ cm$^{-2}$)&$6.8^{+0.5}_{-0.4}$   &$8.3^{+1.8}_{-1.6}$   &$24\pm4$	 \\[0.2cm]
 	  &$\Gamma$		      &$2.43\pm0.05$	     &$1.84^{+0.07}_{-0.06}$&$2.3\pm0.2$ \\[0.2cm]
          &Power law flux             &$9.8\times10^{-12}$   &$1.9\times10^{-11}$   &\nodata\\[0.2cm]
	  &$\chi^2/\nu$ 	      &30.9/28		     & 79/65		    &187/130\\[0.2cm] \hline
Power law &$N_H$ ($10^{20}$ cm$^{-2}$)&$6.3^{+0.6}_{-0.5}$   &$8.0\pm2.0$           &$32^{+18}_{-8}$ \\[0.2cm]
+ Raymond &$\Gamma$		      &$2.38^{+0.9}_{-0.7}$  &$1.78^{+0.08}_{-0.09}$&$2.2^{+0.3}_{-0.2}$\\[0.2cm]
          &Power law flux             &$9.3\times10^{-12}$   &$2.0\times10^{-11}$   &\nodata\\[0.2cm]
          &kT (keV)                   &$0.8^{+0.5}_{-0.3}$   &$0.9^{+0.5}_{-0.7}$   &$0.21^{+0.06}_{-0.05}$ \\[0.2cm]
          &Raymond flux               &$3.1\times10^{-13}$   &$8.4\times10^{-13}$   &\nodata\\[0.2cm]
	  &$\chi^2/\nu$ 	      &21.4/26  	     & 74.5/63  	    &161/128\\[0.2cm] \hline
Power law &$N_H$ ($10^{20}$ cm$^{-2}$)&$5.1^{+1.1}_{-1.5}$   &$12^{+10}_{-5}$	    &$36^{+30}_{-20}$\\[0.2cm]
+ MCBB    &$\Gamma$		      &$2.1^{+0.3}_{-0.8}$   &$1.7^{+0.2}_{-0.3}$   &$2.0\pm0.3$\\[0.2cm]
          &Power law flux             &$9.1\times10^{-12}$   &$1.9\times10^{-11}$   &\nodata\\[0.2cm]
          &kT (keV)                   &$0.27\pm0.03$         &$0.24^{+0.11}_{-0.08}$&$0.17\pm0.05$\\[0.2cm]
          &R$_{in}\times\sqrt{cos\theta}$ (km)&$2.3^{+1.3}_{-1.0}\times10^{3}$&$5^{+49}_{-2}\times10^{3}$&\nodata\\[0.2cm]
          &MCBB flux                  &$8.6\times10^{-13}$   &$4.9\times10^{-12}$   &\nodata\\[0.2cm]
	  &$\chi^2/\nu$ 	      &18.7/26  	     & 68.3/63		    &162/128\\[0.2cm] \hline
Power law &$N_H$ ($10^{20}$ cm$^{-2}$)&$7.1^{+0.7}_{-1.2}$   &$1.3^{+3.0}_{-1.3}$   &$<2.7$ \\[0.2cm]
* edge	  &$\Gamma$		      &$2.57\pm0.10$	     &$1.77\pm0.06$	    &$2.11^{+0.16}_{-0.14}$\\[0.2cm]
          &Flux                       &$1.0\times10^{-11}$   &$2.0\times10^{-11}$   &\nodata\\[0.2cm]
          &E (keV)		      &$0.41^{+0.08}_{-0.14}$&$0.22^{+0.03}_{-0.06}$&$0.29^{+0.03}_{-0.16}$ \\[0.2cm]
          &$\tau$		      &$0.4^{+1.2}_{-0.2}$   &$6^{+8}_{-3}$         &$>6$\\[0.2cm]
	  &$\chi^2/\nu$ 	      &18.1/26  	     & 69.8/63  	    &158/128\\[0.2cm] \hline
Power law &$N_H$ ($10^{20}$ cm$^{-2}$)&$6.8\pm0.5$           &$5^{+3}_{-5}$          &$1^{+6}_{-1}$ \\[0.2cm]
+ gaussian&$\Gamma$		      &$2.48^{+0.11}_{-0.05}$&$1.77^{+0.09}_{-0.11}$ &$1.86^{+0.19}_{-0.12}$\\[0.2cm]
          &Power law flux             &$9.9\times10^{-12}$   &$2.0\times10^{-11}$    &\nodata\\[0.2cm]
          &E (keV)		      &$0.52^{+0.06}_{-0.20}$&$0.29^{+0.07}_{-0.29}$ &$0.35^{+0.06}_{-0.34}$ \\[0.2cm]
          &$\sigma$ (keV)	      &$<0.29$               &$0.09^{+0.14}_{-0.05}$ &$0.09^{+0.11}_{-0.04}$\\[0.2cm]
          &Eq. width (eV)             &-20.6                 &-255                   &-318 \\[0.2cm]
          &Flux taken by Gauss.       &$-3.0\times10^{-13}$  &$-1.7\times10^{-13}$   &\nodata\\[0.2cm]
 	  &$\chi^2/\nu$ 	      &17.8/25   	     & 70.2/62  	     &104/127\\[0.2cm] \hline
\enddata
\tablecomments{The model derived from \cha data for the galactic emission
(Table~\ref{fit}) is also included in the fit as a fixed component for SAX and 
PSPC data. Quoted $N_H$ are in excess of the Galactic line of sight value
($4.1\times 10^{20}$ cm$^{-2}$). 
Fluxes for each model component (in units of \flux) are calculated 
in the 0.5-2.4 keV 
range and are corrected for cold absorption. Quoted errors ar at 90\% 
confidence level for one interesting parameter.}
\label{fit2} 
\end{deluxetable}
\clearpage

\section{DISCUSSION}
\label{disc}
\subsection{Variability.}
We have detected flux variability on various time scales, from a few hours 
to years (Section~\ref{time}). At least 12 of the light 
curves of the individual PSPC and HRI observations (binned on the GTIs) show 
variability at 
a high confidence level, on timescales of $\sim 1$ day. Since the sampling of 
the ROSAT data is sparse, searches for periodic signals or
characteristic timescales are unreliable. 
The variability appears to be characterized mainly by three fast X-ray shots,
rising and decaying in 2 to 20 days (observations 1, 3, 10) followed by
a slow rise over the next few years, and then becoming irregular after 1997.
\subsubsection{Short timescales.} M81 is part of a sample of LLAGN that are 
observed to be 
less variable, on timescales less than a day, than ``normal'' AGN. Four ASCA 
observations of M81 show that its behaviour is consistent with that of the 
other LLAGN in the sample. \citet{ptak2}
compared the amplitude of the variability (through the ``excess variance''
calculation, see \citealp{nandra}) 
of a sample of LLAGN with that of Seyfert galaxies
examined in \citet{nandra} and found that while the latter show a significant
anti-correlation between the X-ray luminosity and the excess variance, the
LLAGNs are well outside of this trend, as their excess variance is apparently
independent of luminosity and indicative of a much smaller amplitude of 
variation. \citet{ptak2} suggested that this difference in the variability trend
of the Seyfert galaxies and LLAGN could be due to the presence of an ADAF in the
LLAGNs. In an ADAF 
the X-ray radiation comes from a region (the ADAF corona) that is much larger 
than the X-ray emitting region of disks in Seyfert nuclei, leading to
small amplitude variations on longer time scales. \citet{pelle}
suggested that the X-ray spectrum of the nucleus of M81 could indeed be
the signature of an ADAF at work, in an intermediate accretion rate regime,
where the main emitting process is inverse Compton scattering rather than
bremsstrahlung.
 
However, using a sample of six quasars plus the \citealp{nandra} sample of AGN,
\citet{fiore} showed that the excess variance $\sigma^2_{RMS}$ is 
instead correlated with the slope of the spectrum, rather than the luminosity, 
and that flatter spectrum 
sources tend to be less variable on a 2-20 day timescale than
steeper spectrum sources. M81 has a relatively flat spectrum and the excess 
variance from three ASCA observations \citep{ptak2} agrees with the 
\citet{fiore} $\sigma^2_{RMS}$ vs. $\Gamma$ correlation for timescales of 1 day 
or less. However, looking at the whole set of
available observations, we find that 2 to 20 days is the timescale of the two
flare events, and this suggests that M81 could lie well above the \citet{fiore}
correlation. The full M81 dataset also does not fit the \citet{ptak2} 
$\sigma^2_{RMS}$ vs. $L$ and instead shows an amplitude variability consistent 
with normal, non ADAF, AGN. A similar behaviour is observed by \citet{glio} in
NGC~4261, a giant elliptical radio galaxy with a LINER nucleus, that hosts a 
low luminosity AGN: on the base of XMM-Newton light-curve and hardness ratio
evolution, the authors conclude that the X-ray variability is related to the
accretion flow.

If the observed flares are due to a dynamical instability,
so that their timescales correspond to the Keplerian period, we can estimate
the radius R at which the flare is produced, using the time interval between the
flare and the closest observation with lower flux (1P-2P = 2d, 2P-3P = 20d) . 
The Keplerian time is:
\begin{equation}
t_K= \left( \frac{R^3}{GM} \right)^{1/2}=1.4\times10^{-5}\sqrt{\frac{R}{R_S}^3}M/M_{\odot}
\end{equation}
where G is the gravitational constant, $R/R_S$ is the radius in units of 
Schwartzchild radii, M is the central mass and $M_{\odot}$ is the solar mass.
Using the estimated limits of $3\times 10^6M_{\odot}< M<6\times 10^7M_{\odot}$
for the mass of the nuclear black hole in M81 \citep{ho,bower}
we get $35<R/R_S<260$ for the 2d flare and $160<R/R_S<1200$ for the 20d
flare.
\subsubsection{Long timescales.}\label{longts} 
The regularity of the rise from late 1994 to 
1997 is broken by some irregular fluctuations in the last 3 years of 
observations (1997-2000, Figure~\ref{flux}). The slow increasing trend could be
interpreted as a steady increment of the accretion rate. In
the case of M81 this increment must be small enough not
to modify the emitted spectrum significantly. Assuming for the central black
hole a mass of $3\times10^6 M_{\odot}-7\times10^7 M_{\odot}$ \citep{ho,bower}, we get an 
Eddington luminosity of $4.5\times10^{44}-9\times10^{45}$ erg/s. If $10\%$ of 
the luminosity is emitted in X-rays (e.g. \citealp{elvis2})
and produced by accretion with efficiency $\eta=0.06$ \citep{frank}, we
would have $L_{Bol}/L_{Edd}= 1.1\times 10^{-3}-2.2\times 10^{-2}$ and an 
accretion rate $\dot{M}=L_{Bol}/(\eta c^2)=3\times 10^{-5}M_{\odot}/yr$. 
If the change in luminosity during the 1994-1997 steady
rise is entirely due to a change in the accretion rate, this would mean that
$\dot{M}$ has doubled during the same period. 

If the nucleus of M81 hosts an
ADAF (e.g. \citealp{nara}), as suggested by \citet{pelle} based on the lack of
reflection features 
and of the thermal blue bump, the inferred accretion rate would locate it
between the quiescent and the
low state regime, characterised by $\dot{M}/\dot{M}_{EDD}\lesssim0.08$ 
\citep{esin}.
\subsection{The spectrum.}
 
As described in Section~\ref{intro}, there is a general agreement on the
description of the 2.0-10.0 keV continuum of the nucleus of M81 (a power law 
with $\Gamma\simeq1.85$: \citealp{ishi,pelle}), while some variety of results 
still exists with regard to the soft X-ray emission. Using \cha data we were
able to estimate the contribution to the soft emission due to circum-nuclear
point sources and hot ISM, and therefore to derive cleaner nuclear spectra from the
larger beams of the ROSAT/PSPC and SAX/LECS.
\subsubsection{The circum-nuclear emission.} The integrated spectrum of the 
point sources ({\bf a}) is consistent with a power-law plus a soft component,
best-fitted with a MCBB with kT=$0.082\pm0.006$ keV and 
a large gaussian-like feature at $0.84^{+0.04}_{-0.05}$ keV (F-test probability
for the addition of this component lower than $10^{-9}$). The 
power-law is consistent with the typical spectra of low mass X-ray binaries
(LMXB). The presence of a thermal component due to accretion from a disk is 
also common in LMXB, but we would expect a much higher temperature ($\sim 1$ 
keV, e.g. \citealp{frank}). Together with the presence of the Gaussian feature,
this suggest that the the soft spectrum could be quite complex, as it includes 
many sources that may have very different spectra. However, an accurate study of
the point-like sources spectrum is beyond the aim of this work. For more details
on this topic see \citet{swa}.
The spectrum of the diffuse emission ({\bf b}) contains a power-law that is very
similar to spectrum {\bf a} and could be due to unresolved compact sources. 
It also includes a significant contribution from a Raymond-Smith spctrum that 
most probably comes from hot ISM. \citet{boro} find a very similar spectrum 
(with the same ratio of 0.5 between the power-law and the thermal component 
fluxes) when analysing the ROSAT/PSPC spectrum of the unresolved emission in 
the central region (less than 8') of M31. Comparing this spectrum with that of 
the LMXBs in the same region, they conclude that the thermal spectrum
is most probably produced by truly diffuse emission from interstellar gas. 

In their analysis of an XMM/RGS spectrum of the nucleus of M81 extracted from a 
strip 52'' to 75'' wide in the cross dispersion direction, \citet{page} 
report the presence of three different thermal components,
with temperature 0.18, 0.64 and 1.7 keV. They conclude that the first two are
consistent with hot ISM produced by supernova remnants, while the third (and
part of the second) could be produced by LMXB. They measure a flux of $8.7\times
10^{-13}$ \flux~ for the sum of these components. This value is roughly half
the flux of spectrum {\bf c} (the sum of spectra {\bf a} and {\bf b}), but it is
not easily comparable with our value, because of the different conditions of
extraction (imaging in ACIS vs. dispersed spectrum in RGS). 

In both {\bf a} and {\bf b} spectra the N{\sc h} is consistent with the 
Galactic value along the line of sight. This implies that the excess of cold
absorption seen in the nuclear spectra collected with the wide beam instruments 
(see below) is intrinsic to the nucleus. 
\subsubsection{The nuclear emission.} 
In the PSPC, BeppoSAX and \cha-T spectra we detect an excess of N{\sc h} with 
respect to the Galactic line of sight value ($4.1\times 10^{20}$ cm$^{-2}$). 
This excess ranges from 2 to 32$\times 10^{20}$ cm$^{-2}$, depending on the 
model and instrument. This is in agreement with the excess of 2 to 
44$\times 10^{20}$ cm$^{-2}$ found by other instruments ({\it Einstein}, 
\citealp{fab88}, ASCA, \citealp{ishi}; BeppoSAX, \citealp{pelle}, this work; 
PSPC, this work, \citealp{imm}; XMM/RGS, \citealp{page}). The 
\cha circumnuclear spectra are well fitted with the Galactic line of sight 
value, clearly showing that this excess of N{\sc
h} is intrinsic to the nucleus.

The spectrum is not consistent with a single
power-law in any of the data sets, all of which show an 
absorption-like feature at $\lesssim0.4$ keV, superimposed on a power-law fit.
A Power-law + MCBB model gives acceptable results, with similar 
values in the ROSAT, BeppoSAX and \cha data. The observed temperature is hotter
than that expected  for a Schwartzchild black hole of the mass of M81
(0.007-0.014 keV) but is consistent with that expected for a Kerr black hole
(0.39-0.83 keV). However, the normalization of the 
model implies an inner disk radius of $<5\times10^4$ km if seen face on. This is
smaller than the  Schwartzchild radius of $(4.5-70)\times10^6$ km for a
$(3-60)\times10^6M_{\odot}$ black hole as in M81. A consistent Power-law+MCBB 
model requires the disk to be almost edge-on ($\theta\sim90^{\circ}$) as well 
as needing a Kerr black hole. \citet{maki} made the
same suggestion in the analogous case of ULXs in galaxies.

Alternatively, the feature can be fitted with an absorption edge.
The optical depth of the edge is variable, and is not seen in all of
the ROSAT data, but its presence and depth are not correlated with the flux.
Also, the observed edge may change energy between the three 
instruments: we find an energy of $0.41^{+0.08}_{-0.14}$ keV in the PSPC,  
$0.22^{+0.03}_{-0.06}$ in BeppoSAX and  $0.29^{+0.03}_{-0.16}$ in \cha.
These measurements are consistent at the 95\% confidence level with the neutral 
C {\sc k} edge at 0.284 keV. Alternatively, we
may be detecting either different spectral features or a single
feature found at different redshifts in different observations. The energy of 
the PSPC feature is consistent with the C{\sc v}  edge at 
0.39 keV and could be produced by an ionized gas in front of the nuclear 
source. 

If the feature seen in
BeppoSAX and \cha is produced by the same ion, this would require the gas to be
substantially redshifted ($z=0.34\pm 0.09$ in \cha-T and $z=0.8\pm 0.4$ in 
BeppoSAX). However, once large redshift are allowed, the PSPC may be
seeing moving matter too, so we cannot exclude that the edge is produced by
other elements (e.g. highly redshifted O{\sc k} edge or O{\sc vii}, or mildly 
ionized Fe, i.e from Fe{\sc ix} at 0.23 keV to Fe{\sc xiv} at 0.39 keV). In any 
case this model suggests that the three
instruments are seeing a dynamically variable system of clouds and that in the
more recent observations the clouds may be accelerating toward the nucleus.
Outflow velocities of 0.1-0.3 $c$ have been reported in several quasars for the
Fe{\sc k} resonant absorption line \citep{chart,reeves}. All of these are
high luminosity objects that are probably radiating at or above their Eddington
limit \citep{kp}, while M81 is radiating at 0.001-0.02 $L_{Edd}$.

A third possibility to fit the absorption features is 
a gaussian absorption line. This model yields by far the best fit to the \cha 
data. The line energy is consistent either with the C{\sc v} K$\alpha$ line at 
0.308 keV or with the C{\sc vi} K$\alpha$ line at 0.367 in all three 
instruments, but again the lack of any feature that could be 
associated with O 
makes this interpretation doubtful. The O{\sc vii} line at 
0.57 keV would be consistent with the best fit energy in the PSPC, but it would 
require a significant redshift to explain the feature in the other instruments
(z$\sim 0.9$ in BeppoSAX and z$\sim0.6$ in \cha), implying a very dynamic
evolution. In either case, however, the observed equivalent width (20 eV in the 
PSPC, 255 eV in
SAX/LECS and 318 eV in \cha/ACIS) would require equivalent hydrogen absorbing
columns larger than $10^{22}$ cm$^{-2}$ \citep{nica} and deep absorption edges
for the corresponding ions (O{\sc vii} at 0.74 keV, C{\sc v} at 0.39 keV and 
C{\sc vi} at 0.49 keV). The lack of these features seems to rule out this 
interpretation.

The main problem with the interpretation of the feature as a C edge
(either neutral or ionized) is that there is no evidence 
of the corresponding O K edge or any other of the edges, expected
between 0.5 and 0.8 keV, which we would expect to see in a thermal gas with
solar composition. The depth of the edge would then require an over-abundance
of C with respect to O by a factor of 10 or more. If the O were depleted onto
large dust grains, the O edge could be substantially suppressed. \citet{gaskell}
argue that large dust grains are common in AGN.
If the number ratio of C/O is above 1, the dust will be rich in oxides, with
almost non-existent carbon grains, leaving an overabundance of atomic C in the
gas phase \citep{whit}. 
\citet{elvis3} suggest that the formation of dust could be triggered by
the free expansion of quasar broad emission line clouds in an outflowing wind. 

The strong flux below the absorption features in both the edge and 
gaussian models strongly suggest the presence of warm gas ($\sim 10^6$K). 
This gas would be in addition to the more highly ionized gas
inferred by \citet{pelle} from the
presence of highly ionized Fe features (the emission line at 6.7 keV and the
absorption edge at 8.6 keV). Dusty warm absorbers have previously been 
suggested in other AGN \citep{komo1,komo2}.

\section{Conclusion}
We have analyzed a large set of observations of the nucleus of M81, including
ROSAT, ASCA, BeppoSAX and \cha observations, in order to study the spectrum and 
variability of this bright source.  

The 20 yr coverage of the nucleus of M81, with different X-ray observatories
(EXOSAT, {\it Einstein}, ROSAT, BeppoSAX, \cha, XMM-Newton) shows variability of
this source on different timescales. In particular, a steady increase of the
nuclear luminosity is suggested by the frequent coverage between years 1990 and
2000, interrupted by shorter flares lasting from 2 to 20 days. Also, flickering
on timescales $< 1$ day is visible in the ROSAT data.

The steady increase of the luminosity can be interpreted as due to a change in
the accretion rate onto the central black hole from the low luminosity (PSPC) 
to the high luminosity (\cha) observations. The 2-20 days timescale of the two 
major flares (observations 1P and 3P) suggest dynamical
instabilities occurring at between 35 to 1200 $R_S$.

Analysis of the ROSAT/PSPC spectra suggests spectral variability, which can be
understood with the variability of a negative residual at energies below 1
keV, relative to the best fit power-law model. This feature is confirmed by an
independent analysis of BeppoSAX and \cha-T spectra.
Spectral analysis with a variety of models (and including the best fit estimate
of the circum-nuclear emission from \cha), suggests that the soft spectra can 
be interpreted either as thermal emission from an highly edge-on accretion disk 
feeding a  Kerr black hole or, even better, as the absorption by warm, 
Oxygen-depleted, gas in front of the nucleus. In the 
absorber case, the gas may be either highly ionized, and in which case it 
must be subject to strong dynamics, or, more likely, low ionization 
low velocity gas.
A physically plausible model to explain the lack of oxygen features is that 
large O-rich dust grains have been formed out of the absorbing gas.

We aknowledge S. Dyson for running the original {\sc galpipe} data reduction for
M81. This research has made use of the High Energy Astrophysics Science Archive 
Research Center (HEASARC), provided by NASA's Goddard Space Flight Center. This 
work was supported by MIUR and by  NASA contract NAS 8-39073 (CXC).


\begin{thebibliography}{}
\bibitem[Arnaud, George and Tennant(1992)]{arn} Arnaud, K. A., 
         George, I. M., Tennant, A. F. 1992, Legacy, 2, 65
\bibitem[Barr et al (1985)]{barr} Barr, P., Giommi, P., Wamsteker, W., 
        Gilmozzi, R., Mushotzky, R. F. 1985, BAAS, 17, 608
\bibitem[Bietenholz et al.(1996)]{biet} Bietenholz, M.F. et al. 1996, \apj,
        457, 604 
\bibitem[Boese(2000)]{boese} Boese, F.G. 2000, A\&AS, 141, 507
\bibitem[Borozdin \& Priedhorsky(2000)]{boro}Borozdin, K.N., \& Priedhorsky, 
        W.C., 2000, \apj, 542, L13 
\bibitem[Bower et al.(2000)]{bower} Bower et al. 2000, \baas, 197, 92.03
\bibitem[Chartas, Brandt \& Gallagher(2003)]{chart} Chartas G., Brandt W.N.,
        Gallagher S.C., 2003, astro-ph/0306125
\bibitem[Conover(1971)]{cono} Conover, W.J. 1971, {\it Practical Nonparametric 
        Statistics}, (Wiley)
\bibitem[Damiani et al(1997)]{dami} Damiani, F., Maggio, A., Micela, G., 
        Sciortino, S 1997, \apj, 483, 370
\bibitem[David et al.(1996)]{david} David, L.P. et al. 1996, The ROSAT Users 
        Handbook, ed. U.G. Briel et al.
\bibitem[de Vaucouleurs et al.(1991)]{rc3} de Vaucouleurs G., de Vaucouleurs A.,
        Corwin H.G., Buta R.J., Paturel G., Fouqu\'e P., 1991, Springer-Verlag, 
	New-York
\bibitem[Elvis \& van Speybroeck(1982)]{elvis} Elvis, M., van Speybroeck, L., 
        1982, \apj, 257, L51
\bibitem[Elvis et al.(1994)]{elvis2} Elvis M., Wilkes B.J., McDowell, J.C.,
        Green R.F., Bechtold J., Willner S.P., Oey M.S., Polomski E., Cutri R., 
	1994, \apjs, 95, 1
\bibitem[Elvis, Marengo \& Karovska(2002)]{elvis3} Elvis M., Marengo M.,
        Karovska M., \apj, 567, L107
\bibitem[Esin, Mc Clintock and Narayan(1997)]{esin} Esin, A.A., Mc Clintock, 
        J.E, Narayan R. 1997, \apj, 489, 865
\bibitem[Fabbiano(1988)]{fab88} Fabbiano, G. 1988, \apj, 325, 544 
\bibitem[Fiore et al.(1998)]{fiore} Fiore, F., Laor, A., Elvis, M., Nicastro,
        F., Giallongo, E., 1998, \apj, 503, 607
\bibitem[Frank, King \& Raine(1992)]{frank} Frank, J., King, A. \& Raine, D.,
        1992, in {\it Accretion Power in Astrophysics}, Cambridge University 
	Press
\bibitem[Freedman et al.(1994)]{fre} Freedman W. L. et al. 1994, \apj, 427, 628
\bibitem[Gaskell et al.(2003)]{gaskell} Gaskell C.M., Goosmann R.W., Antonucci
        R.R.J., Whysong D., 2003, \apj, submitted 
\bibitem[Gliozzi, Sambruna \& Brandt(2003)]{glio} Gliozzi M., Sambruna R.M.,
        Brandt W.N., 2003, \aap, 408, 949
\bibitem[Ho, Filippenko, Sargent(1996)]{ho} Ho, L. C., Filippenko, A. V., 
        Sargent, W. L. W. 1996, \apj, 462, 183
\bibitem[Ho et al.(2001)]{ho4} Ho, L. C., et al. 2001, \apj, 549L, 51
\bibitem[Immler \& Wang(2001)]{imm} Immler, S. \& Wang, Q.D., 2001, \apj, 554
        202
\bibitem[Ishisaki et al.(1996)]{ishi} Ishisaki Y. et al. 1996, \pasj, 48, 237
\bibitem[Iwasawa, Fabian \& Nandra(1999)]{iwa} Iwasawa, K., Fabian A., \&
       Nandra, K., 1999, \mnras, 307, 611
\bibitem[Iyomoto \& Makishima(2001)]{iyo} Iyomoto, N. \& Makishima, K. 2001, 
        \mnras, 321, 767
\bibitem[King \& Pounds(2003)]{kp} King, A. \& Pounds, K. 2003, 
        \mnras, in press
\bibitem[Komossa \& Bade(1998)]{komo1} Komossa S., Bade N., 1998, \aap, 331, L49
\bibitem[Komossa \&  Breitschwerdt(2000)]{komo2} Komossa S., Breitschwerdt D., 
        1998, \apss, 272, 299
\bibitem[La Parola et al.(2001)]{lapa} La Parola, V., Peres G., Fabbiano G., 
        Kim D. W., Bocchino F., 2001, \apj, 556, 47
\bibitem[Makishima et al.(2000)]{maki} Makishima, K. et al. 2000, \apj, 535, 632
\bibitem[Nandra et al.(1996)]{nandra} Nandra, K., George, I.M., Mushotzky, R.F.,
        Turner, T.J., Yaqoob, T., 1996, \apj, 476, 70
\bibitem[Narayan \& Yi(1994)]{nara} Narayan R. \& Yi I., 1994, \apj, 428,13
\bibitem[Nicastro, Fiore \& Matt(1999)]{nica} Nicastro F., Fiore F., Matt G.,
        1999, \apj, 517,108
\bibitem[Page et al.(2003)]{page} Page, M.J., et al.,
        2003, \aap, 400, 145
\bibitem[Peimbert \& Torres-Peimbert(1981)]{pei} Peimbert, M. \& 
         Torres-Peimbert, S. 1981, \apj, 245, 845
\bibitem[Pellegrini et al.(2000)]{pelle} Pellegrini, S., Cappi, M., Bassani, L.,
        Malaguti, G., Palumbo, G. G. C., Persic, M. 2000, \aap, 353, 447
\bibitem[Petre et al.(1993)]{petre} Petre, R., Mushotzky, R.F., Serlemitsos,
        P.J., Marshall, F.E., 1993, \apj, 418, 644
\bibitem[Pfeffermann et al.(1987)]{pfef} Pfeffermann, E. et al. 1987, 
        Proc. SPIE, 733 (519), 117
\bibitem[Ptak et al.(1998)]{ptak2} Ptak, A., Yaqoob, T., Mushotzky, R., 
        Serlemitsos, P., Griffiths, R. 1998, \apj, 501, L37 
\bibitem[Reeves, O'Brien \& Ward (2003)]{reeves}  Reeves J.N., O'Brien P.T., 
        Ward M.J., 2003, ApJ, 593, L65 
\bibitem[Swartz et al.(2003)]{swa} Swartz D.A., Ghosh K.K., McCollough M.L.,
        Pannuti  T.G., Tennant A.F., Wu  K., 2003, \apjs, 144, 213
\bibitem[Tody(1986)]{tody} Tody, D. 1986, SPIE 627, 733
\bibitem[Whittet(2003)]{whit} Whittet D.C.B., {\it Dust in the galactic
        environment}, Bristol: IoP, 2003
\bibitem[Worrall et al.(1992)]{wor} Worrall, D.M. et al. 1992, in Data analysis 
        in Astronomy - IV 
\bibitem[Zimmermann et al.(1994)]{zim} Zimmermann, H. U., Lewin, W., 
        Predehl, P. et al. 1994, Nature, 367, 621
\end{thebibliography}
\end{document}